Acousto-electric Characteristics of Periodically Poled Ferroelectric Plate.

I.V. Ostrovskii, A.B. Nadtochiy

*Department of Physics and Astronomy, University of Mississippi, University, MS 38677*

*Email: iostrov@phy.olemiss.edu*

➢ **Abstract.**

A multidomain two-dimensional periodically poled ferroelectric plate vibrator is reported for the first time. The theoretical calculations, computer simulations by the Finite Element Method and experimental data from the lithium tantalite samples reveal a domain acousto-electric resonance. A polarization inversion in a *y*-rotated cut of a ferroelectric chip is firstly done. The acousto-electric characteristics of the vibrator are calculated and measured.

➢ **Introduction.**

Multidomain ferroelectrics are the subjects of the rigorous fundamental physical investigations and possible applications during last decades since the first publications on fabrication of the periodic inversely poled structures [1-5]. Main areas of research are the fabrication of inversely poled micro-domains in bulk ferroelectric crystals, which are sometime referred to as one dimensional acoustic super-lattice, and their applications for ultrasonic transducers [6-12]. Different oxygen-octahedral ferroelectric crystals ($LiNbO_3$, $LiTaO_3$, $KTiOPO_4$, and $BaTiO_3$) with the periodic domain structures have been reported [1, 3, 4, 9, 12, 13] in a one-dimensional bulk media. Scanning Electron Microscopy technique was employed to show a significant interaction of the surface acoustic waves propagating in a periodically poled lithium niobate [14], which results in some nonlinear effects [15]. The domain walls and associated complexes of crystal lattice defects are reported [16] to be responsible for a new type of nonlinear ultrasonic attenuation in lithium niobate. Last years a fabrication of the ferroelectric domains at the nanoscale is of great interest [17] view the potential applications in information storage devices and ferroelectric memory cells. The optical properties and generation of optical harmonics in periodic ferroelectric structures is another important application. At this point we want to mention a new type of polariton [18] discovered in a piezoelectric superlattice [1, 19].

In point of fact, we can not refer to any publication on acoustoelectric properties of multidomain ferroelectric structures in two-dimensional ferroelectric resonators.

However ferroelectric chips are widely used for fabrication of the surface acoustic wave filters, which are important component of mobile and telecommunication systems.

Our work is devoted to theoretical calculations, computer simulations and experimental verification of the acousto-electric properties of multidomain periodically poled ferroelectric domain structures in the thin plates and resonators made of 3m-symmetry materials like lithium tantalite and lithium niobate.

> **Multidomain periodically poled ferroelectric plate (theory).**

As a typical model, we consider a ferroelectric plate consisting of N pairs of inversely poled domains as shown in the figure 1. Axis Y or X is parallel to wave vector ***k***, and Z-axis is parallel to a vector-normal ***n***. Note an acoustic waves/vibrations propagating along Y and X axes in a Z-cut crystalline plate of a 3m-symmetry crystal have close acoustic properties. For example, the zero symmetrical $S_0$-modes of Lamb wave are practically identical. They both have a longitudinal acoustic displacement accompanying by a longitudinal piezoelectric field, non-dispersive frequency range and a phase speed near ~ $6 \cdot 10^5$ cm/sec. The theoretical and computer calculations for these both cases give practically the same results.

Inverse polarization is shown by the oppositely oriented arrows (+Z and –Z) in fig. 1. Mathematically this configuration is represented by opposite signs of piezoelectric coefficients in the adjacent domains: +*e* for A-type domains and –*e* for B-type domains. This reflects a fact that the same type of mechanical strain generates opposite piezoelectric fields in A- and B-domains.

Equations of motion (1) and electrodynamics (2) describe the vibrations in the electrostatic approximation:

$$\rho \frac{\partial^2 u_i}{\partial t^2} = \frac{\partial T_{ij}}{\partial x_j} \qquad (1)$$

Equation of Poisson: $\quad \dfrac{\partial D_i}{\partial x_i} = 0 .\qquad (2)$

Piezoelectric properties of a plate material are given by the equations (3) and (4).

$$T_{ij} = c_{ijkl}^E \frac{\partial u_k}{\partial x_l} + e_{mij} \frac{\partial \varphi}{\partial x_m} , \qquad (3)$$

$$D_i = e_{ikl} \frac{\partial u_k}{\partial x_l} - \varepsilon_{ij}^S \frac{\partial \varphi}{\partial x_j} , \qquad (4)$$

where, $\rho$ is a density, $u_i$ are the acoustic displacement components, $T_{ij}$ are the mechanical stress tensor components, $\varphi$ is an electric potential, $D_i$ is an electric displacement, $c_{ijkl}^E$ are the elastic modules tensor components at a constant electric field, $e_{mij}$ are the piezoelectric coefficient tensor components, $\varepsilon_{ij}^S$ are the dielectric permittivity tensor components at a constant strain. We omit below the superscripts $E$ and $S$ at the elastic modules and permittivity. The boundary conditions of zero acoustic stress at the plate faces should be satisfied as well: $T_{ij} n_j = 0$. The vectors $n_j$ are the unit vectors normal to the plate surfaces.

In an isotropic plate, two families of acoustic waves propagate: Lambs' waves with acoustic displacements $u_y$ and $u_z$ along $y$ and $z$ axes, respectively and shear normal waves with the displacements $u_x$ along $x$ axis [19]. Standing plate wave is a normal vibration, which satisfies both the equations of motion and the boundary conditions at $z = \pm t$, where $t$ is a half thickness of the vibrator. Plate waves in periodically poled ferroelectrics have not been considered yet. We made the experimental measurements with the Lamb waves in a multidomain sample of lithium tantalite and observed basically same acoustic modes, which exist in a regular single crystal plate [21, 22]. The zero symmetrical mode has phase speed $V = 5.9 \cdot 10^5$ cm/sec. One can also use the Finite Element Analysis (FEM) and make a computer simulation for two dimensional resonators. In the next section, we describe our results on FEM-simulations of the acoustic multidomain resonator.

For a purpose of simplification, we consider an acoustic mode with a negligible dispersion in a certain frequency range. It may be the shear symmetrical zero mode $SS_0$ with $u_x \neq 0$ or Lamb symmetrical zero mode ($S_0$) at low frequency limit when a wave length is much longer than plate thickness and then mainly only $u_y \neq 0$. For a standard 0.35 mm thick lithium tantalite YZ-cut-chip, this frequency range is from zero to approximately 5 MHz.

We consider $S_0$ mode with longitudinal displacements $u = u_y \exp(i\omega t)$ along $y$-axis. The piezoelectric field accompanying wave is $E_y = E = -\partial \varphi / \partial y$, where $\omega$ is an angular frequency. The solutions of the equations (1) – (4) for the acoustic displacement $U$ and electric field $E$ can be derived as:

$$U = \frac{-1}{\rho\omega^2}\frac{\partial T_N}{\partial y} \tag{5}$$

$$E = -\frac{e}{\varepsilon c}T_N + \frac{1-K^2}{\varepsilon}D_N(\omega), \tag{6}$$

where the $T_N$ and $D_N$ are the acoustic stress and electric inductance along the N-pair-domain vibrator, $e$ is a piezoelectric coefficient, $c$ is an elastic module at a constant electric inductance, $\varepsilon$ is a dielectric permittivity at a constant strain, $K^2 = e^2/(c\varepsilon)$ - the squared electromechanical coupling coefficient. The $T_N$ can be derived by solving the differential equation resulting from the system (1 – 4).

$$\frac{\partial^2 T_N}{\partial^2 y} + \frac{\rho\omega^2}{c}T_N + \frac{\rho\omega^2 e(y)}{\varepsilon c}D_N = 0 \tag{7}$$

We note a piezoelectric coefficient $e(y)$ in the equation (7) is a square wave function, which changes a sign at each inter-domain boundary. Thus we have four boundary conditions, they are: two conditions at the ends of the multidomain bar $T(y = L/2) = 0$ and $T(y = -L/2) = 0$, and another two internal boundary conditions; they are equalities of the stress tensor and mechanical displacements at the interfaces between the adjacent inversely poled domains. The last two mean a continuity of the acoustical stresses and mechanical displacements at the inter-domain boundaries. After equation (7) is solved for $T_N$ and the $T_N$ is written via the inductance $D_N$, the solution for the stress $T_N$ is substituted to the equation (6). Consequently we find the equation for the electric inductance $D_N(\omega)$ in N-pair-domain vibrator via electric field $E(y)$. An integral of the electric field $E(y)$ over the length of the vibrator is a sum of the particular integrals over each single ferroelectric domain, and it gives an electric potential $\varphi(y)$ at any point $y$ of the vibrator. This sum over a whole length is equal to an applied rf-voltage $V(\omega) = V_0 \exp(i\omega t)$. For the simplest case of only two inversely poled domains (N=1), the electric potential is

$$\varphi_1(y) = V\frac{ky\cos(kd/2) \mp K^2[\sin(k(d-2|y|)/2) - \sin(k|y|/2)]}{kd\cos(kd/2) - 2K^2\sin(kd/2)}, \tag{8}$$

where sign "–" is for any A-domain and "+" is for any B-domain.

The equation (8) in particular allow to experimentally measuring a distribution of the electric potential over a length of a periodically poled acoustic vibrator. It is an important tool for characterization of the multidomain ferroelectric structures. In figure 2, we present a theoretical dependence (plot 1) and experimental measurements (plot 3) for the lithium tantalite sample LT4a3 with the inversely poled domains of 1 mm length each. Experiment and theory are in a good agreement. The sample LT4a3 is a standard 0.35mm chip of Y-rotated cut with Y-axis rotated at 42° clockwise with respect to the normal vector **n** on fig. 1. This cut actually gives an increase in the effective electromechanical coupling coefficient but not alters the main results of the theoretical calculations.

An important acoustoelectric characteristic is an rf-admittance $Y(\omega)$ of the vibrator, since it shows whether or not the vibrator can be used for rf-filtering and in what frequency range. For the case of two inversely poled domains

$$Y_1(\omega) = \frac{i\omega A \varepsilon}{2L} \left[ 1 - \frac{K^2}{(kL/4)} \tan(kL/4) \right]^{-1} , \qquad (9)$$

where $A$ is an electrode area. The equation (9) yields an anti-resonance frequency $\omega_a = (\pi V/d)$ that is twice higher than for a single crystal vibrator of the same length $L$ and same acoustic wave speed $V$ [20]. This result does not differ from the case of bulk periodic acoustic superlattice structures [8].

In the case of multidomain inversely poled ferroelectric structures with N>1, the equations (8, 9) become the functions of the sums of different trigonometric functions representing all the domains in the multidomain structure. However the physical results remain the same. Despite a total length of the vibrator is $L_N = 2dN$, a main resonance in $Y(\omega)$ is observed at the frequency that corresponds to the length $L_1=2d$ of only two inversely poled domains – equation (9).

There is an interesting result. A total length of the vibrator remains the same and equal to the initial length $L_N = 2dN$ at any instant. It is unlike to the case of a single crystal vibrator that changes its length during a vibration process [20]. This peculiarity of multidomain vibrator along with the frequency characteristics may be appealing for rf-filter design especially for super-high-frequencies up to GHz range. We note at the GHz-frequencies the ferroelectric domains must be at the nanoscale.

> **Finite Element Analysis (computer simulation).**

We consider the ultrasonic vibrations in the multidomain Y-rotated-cut of 3m-symmetry-crystalline plate. Crystallographic *y*-axis is 42°-clockwise rotated with respect to vector normal ***n*** in figure 1. This is one of the typical cut for the chips of lithium tantalite and lithium niobate crystals to be used in numerous acoustoelectric, acoustooptic and optoelectronic applications. The normal shear modes with X-displacements are not piezoactive wave, but Lamb's waves are piezoelectrically active. The solutions of the four-equation system (1) - (2) do not depend on *x*, and can be written in a general form as

$$u_i = a_i(y,z)\exp(i\omega t); \varphi = \varphi(y,z)\exp(i\omega t) \qquad (10)$$

where $a_i$ are the complex amplitudes, $i = x, y$ or $z$, and $\omega$ is an angular frequency. Further we substitute the solutions (11) into the equations (1, 2), then use the relations (3, 4) and the 3m-symmetry crystal constant matrices [20, 23]. As a result, we get a system for three unknown components of the acoustic displacement $a_x$, $a_y$, $a_z$ and fourth unknown $\varphi$ for the electric potential. We consider then Lambs' waves as they are piezoelectrically active. The boundary conditions are imposed in a standard way, that is the normal and tangential components of the stress are equal to zero at plate surfaces.

The whole system of the equations for the unknown amplitudes and boundary conditions is numerically solved by the finite element method [24-26]. According to the standard procedure [26], we introduce three special vectors that describe the generalized coordinate $\{a_y, a_z, \varphi\}$, generalized strain $S_G = \left\{ \dfrac{\partial a_y}{\partial y}, \dfrac{\partial a_y}{\partial z}, \dfrac{\partial a_z}{\partial z}, \dfrac{\partial \varphi}{\partial y}, \dfrac{\partial \varphi}{\partial z} \right\}$ and generalized stress $T_G = \{T_{yy}, T_{yz}, T_{zz}, D_y, D_z\}$. Then the equations (3) and (4) can be expressed in terms of generalized strain $S_G$ and generalized stress $T_G$, that is $T_G = [M]S_G$. The material constant matrix $[M]$ is written in a form, which corresponds to the five-component generalized strain $S_G$ and stress $T_G$.

$$[M] = \begin{pmatrix} c_{11} & c_{13} & -c_{14} & e_{22} & e_{31} \\ c_{13} & c_{33} & 0 & 0 & e_{33} \\ -c_{14} & 0 & c_{44} & e_{15} & 0 \\ e_{22} & 0 & e_{15} & -\varepsilon_{11} & 0 \\ e_{31} & e_{33} & 0 & 0 & -\varepsilon_{33} \end{pmatrix}, \qquad (11)$$

We use the transformation matrix [23, 26] to rotate the crystallographic reference frame of *y, z* axes to the ***k*** and ***n*** axes that are related to sample geometry as shown in figure 1.

$$[P] = \begin{pmatrix} \cos^2\alpha & \sin^2\alpha & 2\cos\alpha\sin\alpha & 0 & 0 \\ \sin^2\alpha & \cos^2\alpha & -2\cos\alpha\sin\alpha & 0 & 0 \\ -\cos\alpha\sin\alpha & -\cos\alpha\sin\alpha & \cos^2\alpha - \sin^2\alpha & 0 & 0 \\ 0 & 0 & 0 & \cos\alpha & \sin\alpha \\ 0 & 0 & 0 & -\sin\alpha & \cos\alpha \end{pmatrix}, \quad (12)$$

where $\alpha$ is an angle of rotation of Y-Z-crystallographic reference frame about X-axis, if any. In the case of 42°-Y-rotated cut the angle α is counterclockwise 48° between normal vector $\mathbf{n}$ and Z-axis in figure 1, which corresponds to 42° clockwise between normal vector $\mathbf{n}$ and rotated Y-axis. We note our calculations show that for different rotated cuts of 3m-symmetry crystals including Y- 42°, 36° or 128°, the pure Lamb's waves still exist. The matrices of the material constants (12) in the original reference frame $[M(y,z)]$ and in the rotated frame $[M'(\mathbf{k},\mathbf{n})]$ are related by the matrix [P] and transposed matrix $[P]^T$ as

$$[M'] = [P]^T \cdot [M] \cdot [P]. \quad (13)$$

In the FEM method, we represent a sample cross-section by a collection of triangles with uniform discretization; a triangle mesh covers a whole cross-section. The acoustic displacements and electric potential are approximated by the linear functions. The numerical calculations are made using the uniform mesh of 155 nodes along the plate length and 5 nodes along the plate thickness. An energy loss in the sample is taken into account by an imaginary part of an elastic constant [23]. The magnitude of the imaginary part of the elastic constant in our crystal is calculated to be 0.25% of a real part of the elastic constant. It corresponds to a mechanical quality factor Q = 200 that is measured experimentally from our samples. We consider a case of an applied rf-electric field along the plate length parallel to wave-vector $\mathbf{k}$ in figure 1. The domain structures with different domain numbers are computed, and the resonator rf-admittance is evaluated.

> **Experimental results.**

In our experiments, we use a standard 0.35 mm tick lithium tantalite chip of Y-42°-rotated cut. We make a sequence of the periodically poled domains by applying an electric field to a 1-mm-periodic metal grid deposited on crystal surface. The domain width of 1 mm and sequence period of 2 mm of the periodically poled domains are chosen to match a non-dispersive frequency range for the zero symmetrical Lamb wave in 0.35 mm thick lithium tantalite waveguide. A phase speed is measured to be $5.9 \cdot 10^3 \, m/s$ in the frequency range 1 to 4 MHz. The number of the inversely poled domain pairs N varies from 7 to12 for different samples. The dimensions of the samples depend on domain number, and for the biggest crystal they are 24 x 20 x 0.35 mm.

A polarization inversion is achieved under 34.3 kV/mm electric field at room temperature. An electric switching current through the sample is monitored to control the polarization inversion process. Note in our samples of Y-rotated cut, the switching voltage is about 1.43 times higher than it is necessary for a Z-cut lithium tantalite (24 kV/mm) [17]. This is obviously related to an effective increase in length along Z-axis of a rotated ferroelectric domain. Since the inclination angle of *z*-axis is 48° counterclockwise with respect to the normal vector, a domain is longer along *z*-axis by a factor of sec48° = 1.49 in comparison with *z*-cut plate. This estimated number is in a good agreement with observed increase of 1.43 times. One can not expect a coincidence of these numbers view the complicated processes of domain inversion, growth, and influence of crystal defects.

The fabricated domain structures are further revealed by using a chemical etching (20 min in 1:2 mix of HF:$HNO_3$ at boiling temperature) and subsequent observation in optical polarizing microscope. To detect an electric potential and its distribution over crystal surface, a pick-up electrode (3 in figure 1) is connected to a digital oscilloscope LeCroy 9400. This signal is further processed in a computer along with the data from a two-dimensional computer controlled stage. All experiments are done in a digital computer controlled mode; the measurements are repeated many times with different samples at room temperature. The typical result on electric potential distribution along crystal length is shown in figure 2 by the circles. The displayed region of 2 mm covers two inversely poled domains. A domain wall location is at zero ordinate. Plot 1 in figure 2 is theoretically calculated by equation (8), and plot 3 is a result of computer simulation by FEM-method. All the three plots demonstrate the same behavior. Displayed in figure 2 distribution may be a tool for the inversely poled domains characterization. We observe the significant distortions in the plot 3 when the domains are of different thickness or with irregular uneven sides.

The rf-admittance is measured with the help of spectrum analyzer Advantest-R3131A. The typical experimental result is shown in figure 3 by the circles. It is taken from the sample LT4a3 with 7 pairs of the inversely poled domains. Theoretical FEM simulation gives plot 2. This plot in turn is very close to the calculations by equation (9), which is not shown to make a figure 3 clearer. The main resonance is observed at 2.947 MHz. One can call it "domain resonance" since it does not exist in the samples without multidomain periodically poled structures; no "domain resonance" exists in a single domain crystal. A frequency of the domain resonance does not depend on number N of inversely poled pairs. The plots 1 and 2 in figure 3 have also some small extra resonances at frequencies 2.78 and 3.24 MHz. They are connected to a complicated nature of the acoustic vibrations of two-dimensional resonator as a whole.

The results of FEM calculations of the longitudinal and transversal displacements in two-dimensional multidomain resonator consisting of $N = 5$ inversely poled domain pairs are shown in figures 4 and 5, respectively.

> **Conclusions**

In this work, we consider the acoustoelectric properties of two-dimensional periodically poled multidomain ferroelectric plate for the first time. A polarization inversion in a 42°Y-rotated cut of ferroelectric chip is firstly done.

A distribution of an electrical potential over multidomain plate surface along wave vector $\vec{k}$ is a sum of the linear and trigonometric functions. A measurement of this distribution may be a new experimental tool for characterization of the periodically poled domain structures in ferroelectrics.

Despite a discrete nature of the Finite Element Method, this model is consistent with the experimental results taken from the two-dimensional multidomain ferroelectric crystal.

The experiments with the periodically poled 42°Y-rotated lithium tantalite, theoretical calculations, and Finite Element Method computations reveal an acousto-electric "domain resonance". The domain resonance frequency depends on the dimension and period of the domains and does not depend on number of inversely poled pairs of domains. This resonance may be used for rf-filter design.

In case of the nanoscale domains, the domain resonance stretches into Gigahertz region for $LiNbO_3$ and $LiTaO_3$ crystals.

> **Acknowledgements.**

This work in part is made possible due to research grant "Acousto-electric phenomena in multidomain ferroelectrics", UM, 2004.

> **References.**

1. D. Feng, N.B. Ming, J.F. Hong, Y.S. Yang, J.S. Zhu, Z. Yang, and Y.N. Wang. Appl.Phys.Lett. **37**, 607 (1980).
2. N.B. Ming, J.F. Hong, D. Feng. J. Mater. Sci. **17**, 1663 (1982).
3. A. Feisst, P. Koidl, Appl. Phys. Lett. **47,** 1125 (1985).
4. V.V. Antipov et al. Sov. Phys. Crystallogr. **30**, 428 (1985).
5. K. Nakamura, H. Ando, H. Shmizu. Appl.Phys.Lett. **50**, 1431 (1987).
6. Yong-yuan Zhu and Nai-ben Ming. Appl. Phys.Lett. **53**, 1381 (1988).
7. O. Yu. Serdobolskaya, G.P. Morozova. Ferroelectrics, 1998, V.208-209, p.395-412.
8. Yong-yuan Zhu, Nai-ben Ming. J. Appl. Phys. **72**, 904 (1992).
9. Yong-yuan Zhu and Nai-Ben Ming. Appl. Phys. Lett. **53**, 2278, (1988).
10. Yong-yuan Zhu, Nai-ben Ming. Wen-hua Jiang, Phys. Rev. B. **40**, 8536 (1989).
11. Yong-yuan Zhu, Shi-ning Zhu,Yi-qiang Qin, Nai-ben Ming. J.Appl.Phys.**79**,2221,(1996).
12. Yong-feng Chen, Shi-Ning Zhu, Yong-Yuan Zhu, Nai-ben Ming, Bio-Bing Jin, ri-Xing Wu. Appl. Phys. Lett. **70**, 592 (1997).
13. G. A. V. Golenishchev-Kutuzov,V. A. Golenishchev-Kutuzov, R. I. Kalimullin. Phys. Usp., **43** (7), 647-662 (2000).
14. D.V. Roshchupkin, M. Brunel, IEEE Trans. on UFFC. **41**, 512 (1994).
15. Golenishchev-Kutuzov A., Golenishchev-Kutuzov V., Kalimullin R. and Batanova N. Ferroelectrics. **285**, 321 (2003).
16. Mack A. Breazeale, Igor V. Ostrovskii, Michael S. McPherson. Jour. Appl. Phys. **96**, 2990 (2004).
17. S. Kim, V. Gopalan, K. Kikamura, Y. Furukawa. J. Appl. Phys. **90**, 2949, (2001).
18. Yong-yuan Zhu, Xue-jin Zhang, Yan-qing Lu, Yang-feng Chen, Shi-ning Zhu, and Nai-ben Ming. Phys. Rev. Lett. **90**, 053903 (2003).
19. M.M. Fejer, G.A. Magel, D.H. Jundt, and R.L. Byer. IEEE J. Quant. Elect. **6**, 911 (2000).

20. Mason W.P. Physical acoustics: principles and methods. V. 1, pt.A, ch.3. Academic Press, New York (1964).

21. A.B.Nadtochiy, A. M. Gorb, and O. A. Korotchenkov. Technical Physics, **49**, 447 (2004).

22. I.E. Kuznetsova, B.D. Zaitsev, S.G. Joshi, and I.A. Borodina. IEEE Trans. Ultrason., Ferroelect., Freq. Contr., vol.48, pp.322-328, Jan. 2001.

23. Auld B.A. Acoustic fields and waves in solids. New York, Wiley, 1973.

24. H. Allik and T. J. R. Hughes, Int. J. Num. Meth. Eng. **2**, 151 (1970).

25. Y. Kagava, T. Yamabuchi. IEEE Trans. Sonic and Ultrasonics. SU-21, p. 275, (1974). 26. 26. N.A. Shulga, A.M. Bolkisev. Oscillations of Piesoelectric Bodies, Naukova Dumka, Kiev, (1990).

> **Figures.**

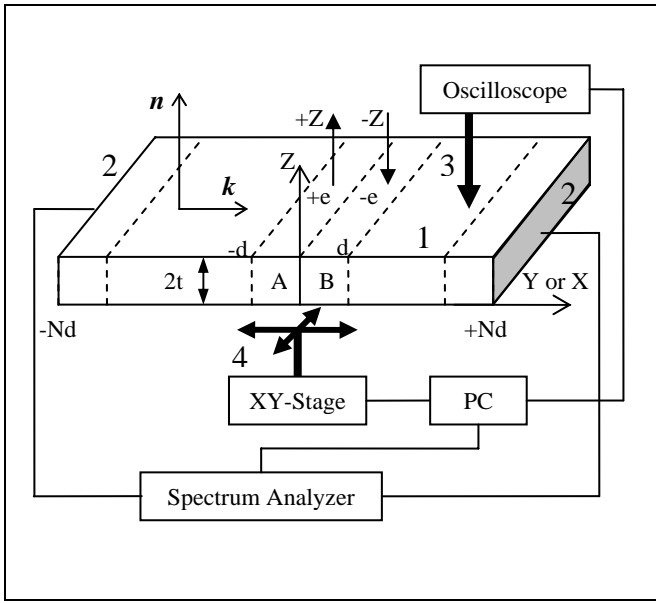

Fig 1. Multidomain ferroelectric plate. 1 – Crystalline plate, 2 – metal electrodes, 3 – pick-up electrode(s) to read the electric potential and acoustic amplitude, 4 – XY stage.

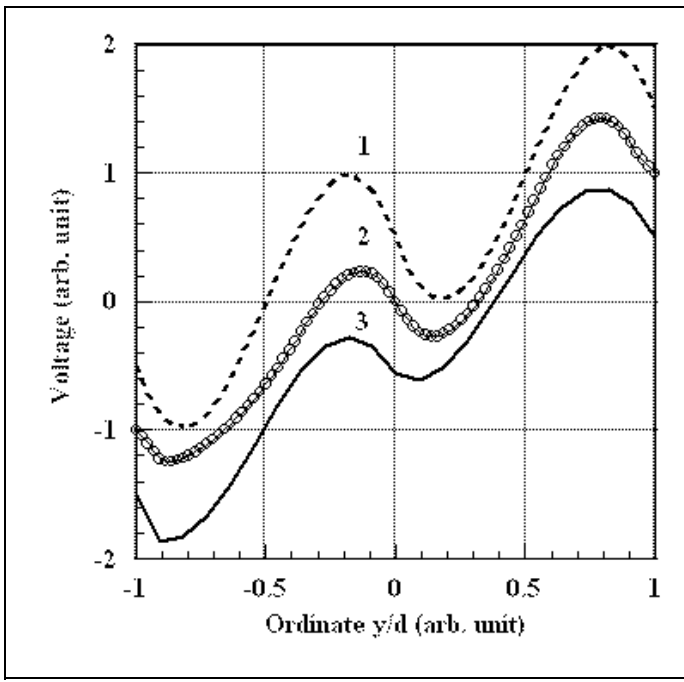

Fig. 2. Distribution of the electric potential along two adjacent inversely poled domains.
Plot 1– calculation by equation (8).
Plot 2 – experimental data from the sample LTa3 at 2.947 MHz.
Plot 3 – FEM computation for the sample LTa3 at 2.947 MHz.

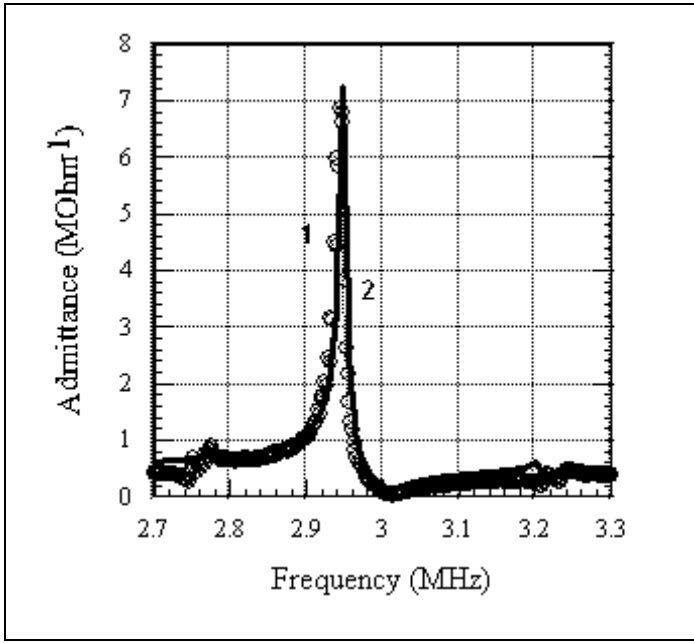

Fig. 3. Plate rf-admittance versus frequency. Plot 1 (circles) – experimental data from the sample LT4a3.
Plot 2 – FEM computations for two-dimensional multidomain periodically poled lithium tantalite crystal with N=7 that corresponds to the sample LT4a3.

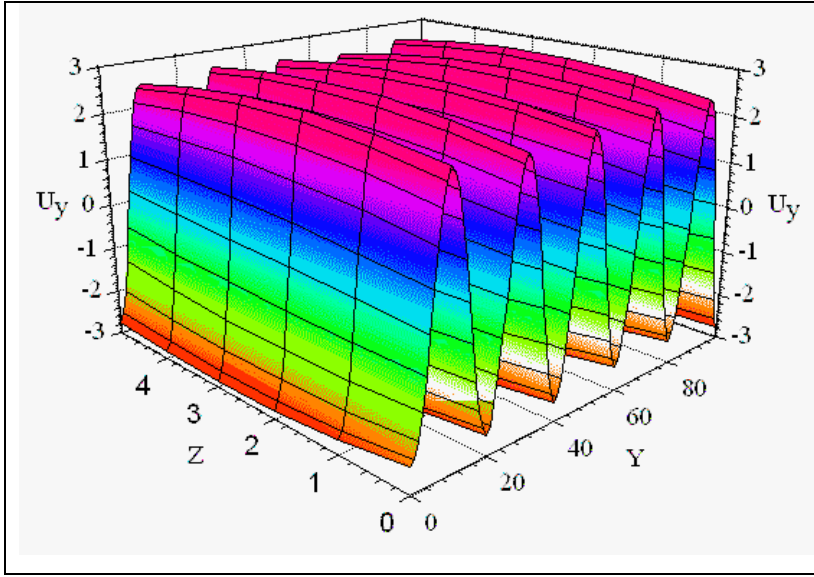

Fig. 4. FEM computation of a tangential displacement $U_Y$ in 2-D periodically poled ferroelectric plate at domain resonance frequency. The calculations are for ZY-cut Lithium Tantalite plate with N=5, a mesh is 101 x 6.
Two plate surfaces are at $Z = 0$ and $Z = 5$.

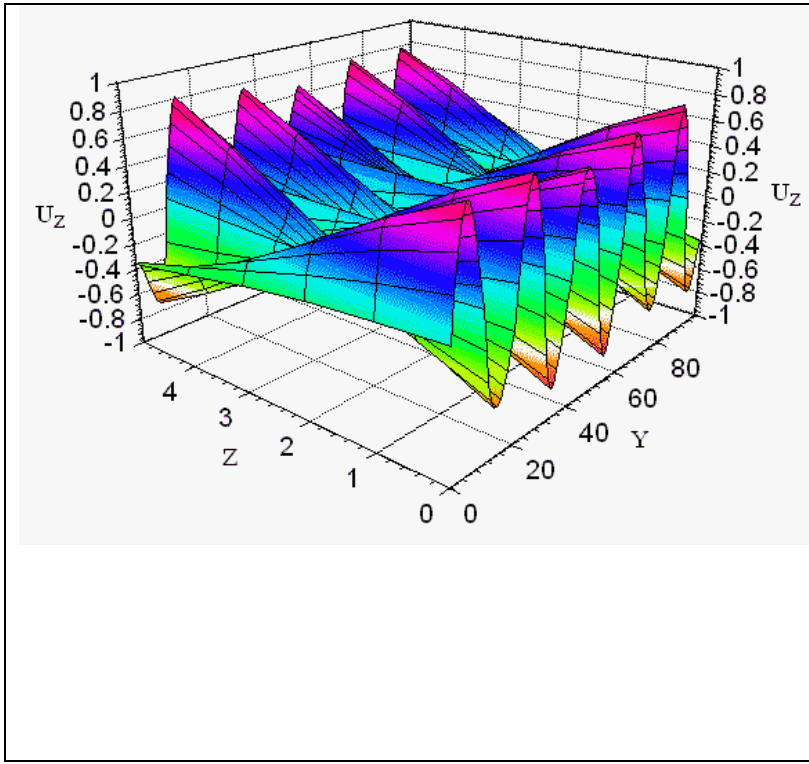

Fig. 5. FEM computation of a normal displacement $U_Z$ in 2-D periodically poled ferroelectric plate at domain resonance frequency. The calculations are for ZY-cut Lithium Tantalite plate with N=5, a mesh is 101 x 6. Two plate surfaces are at $Z = 0$ and $Z = 5$.